\documentclass[aps,pre,superscriptaddress,groupedaddress]{revtex4}
\usepackage{pifont}
\usepackage{array}
\usepackage{stackengine}
\usepackage{graphicx}
\usepackage{subfigure}
\usepackage{multirow}
\usepackage{footnote}
\usepackage{caption}
\usepackage{color}
\usepackage{epstopdf}
\usepackage{chemarr}
\usepackage{amsmath}
\usepackage{dcolumn}
\usepackage{bm}
\newcolumntype{L}[1]{>{\raggedright\let\newline\\\arraybackslash\hspace{0pt}}m{#1}}
\newcolumntype{C}[1]{>{\centering\let\newline\\\arraybackslash\hspace{0pt}}m{#1}}
\newcolumntype{R}[1]{>{\raggedleft\let\newline\\\arraybackslash\hspace{0pt}}m{#1}}
\def\be{\begin{equation}}
\def\ee{\end{equation}}
\def\bea{\begin{eqnarray}}
\def\eea{\end{eqnarray}}

\def\kpe{k_e^+}
\def\kme{k_e^-}
\def\kpf{k_f^+}
\def\kmf{k_f^-}
\def\kpb{k_b^+}
\def\kmb{k_b^-}
\def\mtot{m_{\mathrm{tot}}}
\def\tlag{\tau_{\mathrm{lag}}}
\def\thalf{\tau_{\mathrm{1/2}}}
\def\kapp{k_{\mathrm{app}}}

\begin{document}

\preprint{APS/123-QED}

\title{On the Kinetics of Body \textit{versus} End Evaporation and Addition of Supramolecular Polymers}

\author{Nitin S. Tiwari}
\affiliation{Group Theory of Polymers and Soft Matter, Eindhoven University of Technology,
P.O. Box 513, 5600 MB Eindhoven, The Netherlands}
  \author{Paul van der Schoot}
\affiliation{Group Theory of Polymers and Soft Matter, Eindhoven
University of Technology, P.O. Box 513, 5600 MB Eindhoven, The Netherlands}
\affiliation{Institute for Theoretical Physics, Utrecht University,
Leuvenlaan 4, 3584 CE Utrecht, The Netherlands}

\date{\today}

\begin{abstract}
The kinetics of the self-assembly of supramolecular polymers is dictated by how monomers, dimers, trimers etc., attach to and detach from each other. It is for this reasons that researchers have proposed a plethora of pathways to explain the kinetics of various self-assembling supramolecules, including sulfur, linear micelles, living polymers and protein fibrils. Recent observations hint at the importance of a hitherto ignored molecular aggregation pathway that we refer to as ``\textit{body evaporation and addition}''. In this pathway, monomers can enter at or dissociate from any point along the backbone of the polymer. In this paper, we compare predictions for the well-established \textit{end evaporation and addition} pathway with those that we obtained for the newly proposed \textit{body evaporation and addition} model. We quantify the lag time, characteristic of nucleated reversible polymerisation, in terms of the time it takes to obtain half of the steady-state polymerised fraction and the apparent growth rate at that point, and obtain power laws for both as a function of the total monomer concentration. We find, perhaps not entirely unexpectedly, that the \textit{body evaporation and addition} pathway speeds up the relaxation of the polymerised monomeric mass relative to that of the \textit{end evaporation and addition}. However, the presence of the \textit{body evaporation and addition} pathway does not affect the dependence of the lag time on the total monomer concentration and it remains the same as that for the case of \textit{end evaporation and addition}. The scaling of the lag time with the forward rate is different for the two models, suggesting that they may be distinguished experimentally.
\end{abstract}

\pacs{Valid PACS appear here}
\maketitle

\section{Introduction} \label{sec1}
Supramolecular polymerisation processes are of immense importance in biology and in chemistry. \cite{tomdegreef} Some classic examples in biology include actin and microtubule self-assembly that play important roles in the context of the mechanics of the cell, and $\beta-$amyloid and prion protein aggregation implicated in neuro-degenerative diseases. \cite{takalo, blanchoin, fletcher, chiti, murphy} Similarly, supramolecular polymerisation has major applications in the chemistry of medicine and in molecular electronics. \cite{aida} In this light, it is not surprising that researchers have long studied the thermodynamic and kinetics of self-assembly. As the time evolution of self-assembly is very much system specific, in particular the early time kinetics, itself the most extensively studied aspect of reversible polymerisation, a whole host of molecular pathways of supramolecular self-assembly have been proposed. \cite{oosawa, cates_all, cohen}

It is generally believed that Oosawa was the first to suggest a model in the context of the polymerisation of actin filaments, where a monomer can be added to or removed from the ends. \cite{oosawa, cohen} Oosawa's model has one important ingredient, known as \textit{nucleation}. This means that a stable critical nucleus of $n_c \ge 1$ monomers must be formed before polymer growth commences. Although Oosawa's model of self-assembly is in agreement with experimental data in the context of actin polymerisation, it fails to describe many other protein aggregation processes. \cite{curve_fitting} Indeed, the prevalent molecular pathway for self-assembly is dictated by the complex molecular structure of the monomers involved, as well as by the type of bonding between monomers that form a polymer. This results in molecular pathways that are more complex than the simple pathway proposed by Oosawa, which is sometimes also referred to as \textit{end evaporation and addition} \cite{dubbeldam, cates_eea, semenov_eea}.

In the context of the living polymerisation or linear polymers, researchers proposed a plethora of pathways by which self-assembly can occur, e.g., polymer \textit{scission and recombination} \cite{cates_sr1,semenov_sr,cates_sr}, \textit{secondary nucleation} \cite{cohen, tuomas_review} and \textit{two-stage nucleation}. \cite{knowles_two_stage} Further work shows that the kinetics of self-assembly is strongly dependent on which of the above mentioned pathways are active in the assembly process. \cite{curve_fitting} The influence molecular aggregation pathways have on the early time kinetics of linear self-assembly, which is the most studied aspect of the problem in hand, motivates researchers to study and characterize all possible pathways. \cite{tuomas_science, cohen, knowles_two_stage, Tiwari_2016, tuomas_review}

With the aim to probe the molecular pathway responsible for linear self-assembly in the context of supramolecular polymerisation, Albertazzi \textit{et al.} recently performed experiments with a self-assembling molecule known as 1,3,5 benzenetricarboxamide or BTA for short. \cite{albertazzi} By imaging the supramolecular polymers at different assembly times, they were able to investigate monomer mixing on the scale of individual polymers. Because their observations could not be explained by any of the hitherto known molecular pathways, they suggested the need to revisit theoretically and experimentally the dynamic behaviour of supramolecular polymers. From their observations, they conclude that the molecular pathway responsible for self-assembly of BTAs is the one in which the monomer can be removed from and inserted anywhere along the polymer backbone.

\begin{figure}[!ht]
\begin{center}
\includegraphics[width=6.5in]{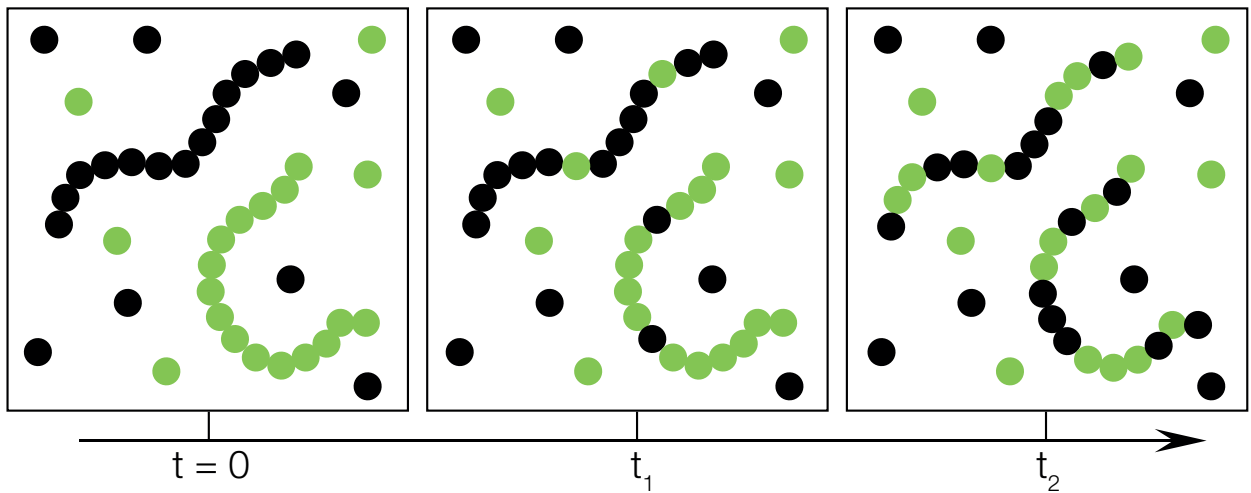}
\end{center}
\caption{Schematic showing the effect of \textit{body evaporation and addition} pathway on the mixing and growth of differently dye-labeled but otherwise identical monomers at time $t=0$. Note the homogeneous insertion of black monomers along the green polymer backbone and \textit{vice versa} via the free monomers in the solution.}
\label{fig1}
\end{figure}

We have schematically depicted the novel pathway in Fig. \ref{fig1}. At time $t=0$ the solution contains only two types of supramolecular homopolymer, and as time progresses mixing of monomers occurs at the supramolecular level. However, contrary to crowding of differently labeled monomers at the ends, which is to be expected if monomers can only attach on and detach from the ends, the mixing takes place homogeneously along the polymer backbone. We call their proposed pathway the ``\textit{body evaporation and addition}'' pathway to contrast it with the conventional \textit{end evaporation and addition} pathway, and study theoretically the kinetics of this pathway in the presence of the \textit{scission and recombination} pathway and \textit{primary nucleation}. \cite{oosawa} The molecular pathway \textit{scission and recombination} allows polymers to break at any point on the backbone resulting into two smaller polymers and recombining two polymers into a longer polymer. \textit{Primary nucleation} is the mechanism by which a critical number of monomers spontaneously self-assemble to form the shortest stable polymer.\cite{oosawa}

Many researchers have concluded that the time evolution of the length distribution of living polymers is typically a mixture of molecular pathways, which are switched on and off dependent on the system of interest. \cite{tuomas_science, hong, hong_moment_closure} It is for this reason that we study the kinetics of the newly proposed pathway in combination with the molecular pathways already referred to. At first glance, \textit{body evaporation and addition} pathway looks similar to \textit{end evaporation and addition}, and naively one would perhaps presume that a simple renormalisation of the rate constants can account for the former. However, a closer look at the problem reveals that for \textit{end evaporation and addition} every polymer has only two ends, resulting into a probability of addition or removal of monomer at the ends that plausibly is independent of the length of the polymer. In contrast, in the case of \textit{body evaporation and addition} the probability of adding or removing a monomer is proportional to the number of bonds in a polymer in which a monomer is being added. Hence, the addition or removal of a monomer along the polymer backbone in the \textit{body evaporation and addition} depends on the size of that particular polymer. As the size of an individual polymer changes as a function of time, it is not possible to simply renormalise the rate constants associated with \textit{body evaporation and addition} pathway and expect it to behave like \textit{end evaporation and addition}.

To study the kinetics of linear self-assembly we start by writing the discrete rate equations. However, the rate equations are highly nonlinear and have so far eluded exact analytical solution except in a few limiting cases. \cite{cohen, hong} Hence, we study the kinetics of self-assembly by closing the discrete reaction rate equations by insisting on plausible approximations. This way, we obtain dynamical equations for the first two moments of the polymer length distribution. These are the number of polymers and the polymerised monomeric mass, of which the latter is primarily probed in assembly experiments. \cite{tuomas_science, hellstrand, tuomas_review} We obtain asymptotic analytical solutions of the resulting dynamical equations. From our analytical solutions we quantify the early time kinetics and show that the scaling of the lag time with the total monomer concentration is identical to that of the standard \textit{end evaporation and addition} pathway even in the presence of the proposed \textit{body evaporation and addition}. However, the lag time significantly does decreases with increasing predominance of \textit{body evaporation and addition} pathway. We also show that when only one of the two addition and evaporation pathways is present, the half-time and the apparent growth rate differ in their scaling with the forward rate constants of monomer insertion.

The remainder of this paper is organised as follows: In Section \ref{sec2}, we introduce the moment equations for the generalized molecular pathway and study the equilibrium properties of the two moments of the length distribution. In Section \ref{sec3}, the dynamical equations for the moments are solved in the absence of polymer recombination and in the limit of vanishing fragmentation and nucleation rate constants. The analytical solutions are then compared with numerical solutions, and used to calculate the half-time and the apparent growth rate for the polymerisation kinetics. Finally, in Section \ref{sec4}, we discuss the results and findings of our theoretical analysis.

\section{Master Equations and Moment Equations} \label{sec2}
The molecular pathways that we include in our study are i) \textit{primary nucleation}, ii) \textit{end evaporation and addition}, iii) \textit{body evaporation and addition} and iv) \textit{scission and recombination}. \cite{cates_all, knowles_two_stage, Tiwari_2016} Our moment equations, derived from the generalized rate equations, allow us to probe the quantitative role of the various molecular pathways involved. The molecular pathway of interest can in principle always be made dominant by switching off other pathways completely or asymptotically. In order to study the kinetics of nucleated polymerisation we first transform the molecular pathways into the corresponding reaction representation.
\begin{align} \label{eq1}
& \text{i) primary nucleation} \hspace{85pt} n_c x \xrightleftharpoons[k_n^-]{k_n^+} y_{n_c},  \\
& \text{ii) end evaporation and addition} \hspace{26pt} y_i + x \xrightleftharpoons[2 k_e^-]{2 k_e^+} y_{i+1}, \\
& \text{iii) body evaporation and addition} \hspace{19pt} y_i + x \xrightleftharpoons[k_b^-(i-1)]{k_b^+(i-1)} y_{i+1}, \\
& \text{iv) scission and recombination} \hspace{40pt} y_i + y_j \xrightleftharpoons[k_f^-]{k_f^+} y_{i+j} \quad \text{for} \quad i,j \le n_c,
\end{align}
where $x$ and $y_i$ are the concentration of monomers and that of polymers of degree of polymerisation $i$, respectively, and $n_c$ is the critical nucleus size, i.e., the degree of polymerisation of smallest stable polymer. Furthermore, $k_n^+$, $k_n^-$, $k_e^+$, $k_e^-$, $k_b^+$, $k_b^-$, $k_f^+$ and $k_f^-$ are the rates of nucleus formation and disintegration (subscript n), monomer addition and removal from the ends (subscript e), the rates of monomer addition and removal from the polymer backbone excluding the ends (subscript b) and polymer recombination and scission (subscript f). The factor of $i-1$ in reaction iii), describing body evaporation and addition, accounts for the fact that a monomer can be added in $i-1$ places on the backbone a polymer of size $i$, and that any one of $i-1$ monomers can be removed from a polymer of size $i+1$ because removal from the ends is forbidden for this pathway.

The reactions are assumed to be reaction limited rather that diffusion limited, implying that the reaction rates
are constant in time. The indices $i$ and $j$ for reaction iv) obey $i,j \ge n_c$, where $n_c \ge 2$. We consider the case $n_c \ge 2$ in order to be able to close the sums in the master equation and obtain the dynamical equations for the first two principal moments of the polymer length distribution. In the case $n_c \ge 2$, a monomer is not counted as a polymer, whereas for $n_c=1$, a monomer can be an active monomer, i.e., a polymer of size one, or an inactive monomer. Hence, the master equations for $n_c \ge 2$ are fundamentally different from that for $n_c=1$ polymerisation.

One additional assumption that we employ in order to close the sums in our master equations is that we assume $k_n^-=k_e^-$. Later we will see that this approximation does not alter our results because in order to be able to close our discrete master equation, we neglect disintegration of a nuclei via the \textit{end evaporation and addition} pathway. Our approximation of irreversible nucleus formation has been employed in the past by several researchers and the results have been quantitatively compared with the experimental data on protein polymerisation, justifying our approximation. \cite{tuomas_science, hong, tuomas_review}

Before we delve deeper into our analysis, a few remarks should be made. In principle, we consider four mechanisms that are responsible for the time evolution of the length distribution: the \textit{primary nucleation}, the \textit{end evaporation and addition}, the \textit{body evaporation and addition} and the \textit{scission and recombination} pathway. Our goal in this work is very specific and is to compare the \textit{early time} kinetics of the two pathways of interest, which are the \textit{end evaporation and addition} and the \textit{body evaporation and addition}. However, if we do not include the \textit{scission and recombination} pathway, the resulting dynamical equations will be singular, meaning that we can make a parameter corresponding to this pathway small but never put it to zero.  Hence, it is for purely mathematical reasons that we make use of the most general description that formally includes all the aforementioned pathways. This is also the reason why we implement \textit{primary nucleation}. To make the problem mathematically tractable we work in the limit of strongly nucleated systems, where for the time domain of our interest the primary nucleation step is not functional and hence can be ignored for all practical purposes. This said, we will explain in detail all of our approximations and limitations, as and when they come in this paper.

To derive a closed form of the moment equations, we start with the discrete master equation for the reaction schemes defined above. For the polymers, this yields
\bea \label{eq2}
\frac{dy_i(t)}{dt} &=& k_n^+ x(t)^{n_c} \delta_{i,n_c} + 2 \kpe x(t) y_{i-1}(t) - 2 \kpe x(t) y_{i}(t) + 2 \kme y_{i+1}(t) - 2 \kme y_i(t) \nonumber \\
& & + \kpb (i-2) x(t) y_{i-1} - \kpb (i-1) x(t) y_{i} + \kmb (i-1) y_{i+1} - \kmb (i-2) y_i \nonumber \\
& & - \kmf (i- 2 n_c +1) y_i(t) + 2 \kmf \sum_{j=i+n_c}^{\infty} y_j(t) + \kpf \sum_{k+l=i} y_k(t) y_l(t) \nonumber \\
& & - 2 \kpf y_i (t) \sum_{j=n_c}^{\infty} y_j,
\eea
where the first term on the right-hand-side  of Eq. \ref{eq2} accounts for the formation of the critical nucleus, the next four terms stem from the \textit{end evaporation and addition} pathway, and terms six to nine represent the \textit{body evaporation and addition}. The last four terms result from \textit{scission and recombination}. Here, $\delta_{i,n_c}$ denotes Kronecker delta that acquires value of 1 when $i=n_c$ and is zero otherwise. Notice that Eq. (5) is missing the term for nucleus disintegration. This is due to our approximation of $k_n^- = k_e^-$ that allows nucleus to disintegrate via monomer removal from an end hence the nucleus disintegration term gets absorbed in the end evaporation terms.

The factor of $(i- 2 n_c +1)$ in the tenth term on the right-hand side of Eq. \ref{eq2} accounts for the number of bonds allowed to break such that the fragmenting filaments are larger than the nucleus size $n_c$. It should be mentioned that the same term, in principle, should include a factor of $\theta(i-2 n_c)$. However, this factor would prevent us from closing the summations and obtaining the dynamical equations for the first two moments of the full polymer length distribution. Hence, we choose not to include it in our master equation. As a consequence, an inconsistency arises for the dimers and trimers, at least if we focus on the case of $n_c=2$. Indeed, this choice would in principle allow a dimer and a trimer to break via the polymer scission mechanism, yet this is disallowed in our way of implementation of reaction schemes. The reason is that in our final analysis we set the limit of $k_f^- \rightarrow 0$, justifying our approximation. Lastly, in the eleventh term, i.e., polymer scission term, the lower limit $i+n_c$ makes sure that two fragments post-scission are stable polymers of size greater than or equal to $n_c$. The condition of the conservation of mass finally results into a time-dependent equation for the monomers
\bea \label{eq3}
\frac{dx(t)}{dt} = -\frac{d}{dt}\left( \sum_{i=n_c}^{\infty} i y_i(t) \right).
\eea

Eq. \ref{eq2} is different from previously obtained master equations on account of the additional terms that describe the contribution of the body evaporation $\kmb$ and addition $\kpb$. \cite{cohen, hong} We also implement the \textit{scission and recombination} pathway to allow for polymer fragmentation resulting only into fragments greater than or equal to the critical size; the recombining polymers also have to be of the size $i \ge n_c$. The reason for this is that we assume that any fragment of size $l \le n_c$ is highly unstable and quickly disintegrates to $l$ monomers. This prohibits the recombination of fragments smaller than the critical nucleus, as they do not exist in a polymeric state. Additionally, by disallowing the fragmentation that results in a fragment smaller than the critical size we prevent this step from contributing to the free monomeric pool. As a consequence, we completely decouple the \textit{end evaporation and addition} from the \textit{scission and recombination} pathway, i.e., one of the pathways can be switched on or off without affecting the other. \cite{cohen}

Our master equation as those very much like it discussed at length in the literature are being highly non-linear equations, and have so far eluded exact analytical solution. \cite{tuomas_review} Hence, a standard practice in the field is not to study the full length distribution, but only the first two moments of it. \cite{oosawa, cohen} These are the polymer concentration, $P$, and the polymerised monomeric mass, $M$. The dimensionality of both are in moles per liter if the rate constants are given in molar units. Of these two quantities, the latter quantity is readily measured by means of, e.g., circular dichroism or fluorescence microscopy. \cite{hellstrand, kelly_cd} The number concentration of polymers can in principle be quantified by measuring the mean size of the polymers, using techniques such as static and dynamic light scattering, and calculating the ratio of the polymerised mass to the mean degree of polymerisation. \cite{zhang_rev}

The two principal moments expressed in our variables read
\bea \label{eq4}
P = \sum_{i=n_c}^{\infty} y_i,
\eea
for the polymer concentration and
\bea \label{eq5}
M = \sum_{i=n_c}^{\infty} i y_i,
\eea
for the polymerised mass. We obtain the dynamical equations for $P$ and $M$ by extracting the first two principal moments from the full polymer length distribution described by Eq. \ref{eq2}. In the process of deducing the dynamical equations for the moments, the only approximation we employ is that we neglect all terms arising from the disintegration of nuclei, which for early times are negligible in number anyway, at least in the limit $k_n^+ \rightarrow 0$. See Appendix A for details.

The dynamical equation for the number of polymers $P(t)$ that we obtain reads
\bea \label{eq6}
\frac{dP(t)}{dt} &=& - \kpf P(t)^2 + \kmf \left[ M(t)-(2 n_c -1) P(t) \right] + k_n^+ \left[ \mtot-M(t) \right]^{n_c},
\eea
and for the time evolution of the the polymerised monomeric mass $M(t)$ we find
\bea \label{eq7}
\frac{dM}{dt} &=& 2 \left[ (\mtot-M(t)) \kpe P(t) - \kme P(t) \right] + \kpb (\mtot-M(t)) (M(t)+P(t)) \nonumber \\
& &+ \kmb (2 P(t)-M(t))+ n_c k_n^+ (\mtot-M(t))^{n_c},
\eea
where $\mtot$ is the total concentration of monomers in the system. A detailed derivation of Eqs. \ref{eq6} and \ref{eq7} from Eqs. \ref{eq2} and \ref{eq3} is provided in Appendix A. For this general set of equations the initial polymerised monomeric mass can have any value between and including, 0 and $\mtot$, i.e., $0\le M(0) \le \mtot$. The same holds true for $P(0)$ and $M(t) \ge P(t)$ for all times, where the equality holds only when all the polymers are critical nuclei.

In the absence of the \textit{body evaporation and addition} terms, Eqs. \ref{eq6} and \ref{eq7} have been compared with kinetic Monte Carlo simulations that do allow for the disintegration of nuclei. \cite{Tiwari_2016} As expected, the time evolution of the moments obtained from the simulation are in quantitative agreement with Eqs. \ref{eq6} and \ref{eq7} in the appropriate limit of strongly nucleated polymerisation, justifying our approximation. With regard to the domain of validity of the Eqs. \ref{eq6} and \ref{eq7}, they produce nonzero and positive $M(\infty)$ and $P(\infty)$ only above the so-called critical concentration of monomers. \cite{paul_review} In the absence of other pathways, except \textit{end evaporation and addition}, the critical concentration obtained from our kinetic equations has a simple analogy to that of the thermodynamic theory of linear self-assembly. \cite{paul_review} However, in the presence of more complex pathways, the mapping is not so trivial as the kinetic theory demands the introduction of additional energy scales associated with the various pathways. For example, an energy scale for monomer removal or addition along the backbone of a polymer would be needed to characterise the \textit{body evaporation and addition} in addition to an energy scale associated with the monomer addition to or removal from the ends of a polymer.

Notice that Eqs. \ref{eq6} and \ref{eq7} are highly nonlinear offering little hope of exact analytical solution. This implies that we have not improved the state of affairs significantly in the context of the various  simplifications of the equations that we have already implemented. Of course, one may linearise Eqs. \ref{eq6} and \ref{eq7} and obtain linear solutions, but this is of limited help as a linear solution can never result in sigmoidal kinetics. Hence, with the aim to reduce the nonlinearity of the equations but preserve the most important and generic behaviour of the kinetics of self-assembly, we employ several additional approximations. Our first approximation is to restrict our analysis to the limit $k_n^+ \rightarrow 0$. This way we reduce the degree of the polynomial, which is essential to obtain an analytical solution. In addition, we also neglect the polymer recombination $\kpf P(t)^2$ term in Eq. \ref{eq6}. By doing so, we break the reversibility condition and hence do not expect the system to follow the corresponding law of mass action. However, earlier studies on nucleated self-assembly have shown that in the context of the early time kinetics we focus attention on, polymer recombination does not play a significant role and hence can be ignored as in fact we shall also make plausible below. \cite{knowles_two_stage} Indeed, by identifying the most dominant terms in the dynamical equations for the polymerisation kinetics, the predicted lag phase has been shown to be in a quantitative agreement with theoretical models, at least in the absence of \textit{body evaporation and addition}. \cite{tuomas_science, knowles_two_stage, hong, hong_moment_closure,  subramaniam} This motivates us to do the same for our reaction pathway, despite it potentially being inaccurate in the long-time limit.

The steady-state solution of Eqs. \ref{eq6} and \ref{eq7} sheds some light on the role of molecular pathways and how our approximations, i.e., $k_n^+ \rightarrow 0$ and $k_f^+ P^2=0$, impact upon the long-time behaviour. In the steady state, the time derivatives of the two moments are zero and we are left with algebraic equations. If we define $K_f=\kpf/\kmf$ to be the fragmentation equilibrium constant, we can switch off the effect of \textit{scission and recombination}, at least in equilibrium, by taking the asymptotic limit $K_f=0$. If $k_f^+=0$ this will be true for any value of $k_f^- \neq 0$. Within this approximation we can compare the effect of \textit{end evaporation and addition} and \textit{body evaporation and addition} explicitly. In this limit the equilibrium polymerised monomeric mass for a nucleus size of $n_c=2$ becomes
\bea \label{eq8}
M(\infty) = \mtot -\frac{\kmb+2 \kme}{4 \kpb + 2 \kpe}.
\eea
In the absence of \textit{body evaporation and addition}, i.e., for $\kpb=\kmb=0$, the polymerised monomeric mass becomes $M(\infty)=\mtot-\kme/\kpe=\mtot-K_e^{-1}$, where $K_e \equiv \kpe/\kme$. In the absence of \textit{body evaporation and addition}, the ratio of rate constants $K_e$ can be mapped onto the equilibrium constant used in the thermodynamic theory of linear self-assembly. \cite{paul_review} In that case, in the polymerised regime, $y_1(\infty)= K_e^{-1}$, remains equal to the critical concentration. Hence, in this limit the polymerised monomeric mass is in agreement with the thermodynamic theory of self-assembly. \cite{paul_review} However, in the presence of \textit{body evaporation and addition} pathway the effective elongation constant has to be renormalized to account for the free energy associated with monomer addition to or removal from the ends and that associated with monomer addition or removal along the backbone of the polymer.

In the limit $K_f \rightarrow 0$ the steady-state number of polymers $P$ and average degree of polymerisation $L$ are given by
\bea \label{eq9}
P(\infty) &=& \frac{M(\infty)}{(2 n_c -1)},
\eea
and
\bea \label{eq10}
L(\infty) &=& \frac{M(\infty)}{P(\infty)} = 2 n_c -1
\eea
for all values of $\kpb$ and $\kmb$. Note that the equilibrium average degree of polymerisation $L(\infty)$ only depends on the critical nucleus size, $n_c \ge 2$, not on the concentration. This, of course, disagrees with the thermodynamic theory but, as we shall see below, this does not preclude very large values of $L(t)$ for intermediate times. \cite{paul_review} Similar results were obtained by Cohen \textit{et al.} in the absence of \textit{body evaporation and addition}. \cite{knowles_two_stage}

\section{Lag Time Analysis} \label{sec3}
\begin{figure}[!ht]
\begin{center}
\includegraphics[width=4in]{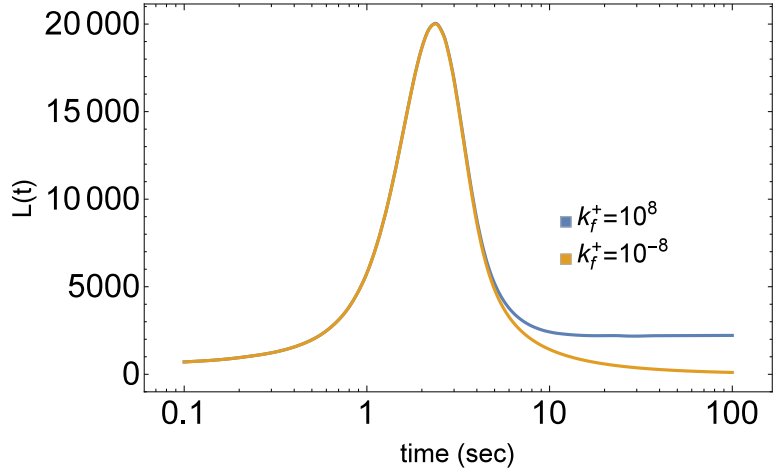}
\end{center}
\caption{Time evolution of the mean length of the polymer $L(t)=M(t)/P(t)$, obtained by numerically solving Eqs. \ref{eq6} and \ref{eq7} for two values of recombination rate $\kpf=10^{8}$ and $10^{-8} M s^{-1}$. The remainder of the system parameters are $k_n^+=10^{-5} s^{-1}$, $\kpe=5 \times 10^{5} M^{-1} s^{-1}$, $\kme=10^{-2} s^{-1}$, $\kpb=5 \times 10^{5} M^{-1} s^{-1}$, $\kmb=10^{-2} s^{-1}$, $\kmf=10^{-4} s^{-1}$, $\mtot=10^{-5} M$, $M(0)=\mtot \times 10^{-4}$ and $P(0)=\mtot \times 10^{-6}$, where $M$ is moles per liter and $s$ is in seconds. The values of the chosen rate constants do not correspond to any particular experiment. However, the order of magnitude is similar to the parameters found in the literature of protein polymerisation. \cite{tuomas_review} The initial conditions are chosen to be small but non-zero, because of the necessity to seed the polymerisation process in the limit of $k_n^+ \rightarrow 0$.}
\label{fig2}
\end{figure}

The steady-state solutions of the moment equations clearly indicate that our generalized reaction schemes do not result into the long polymers in the limit of $t \rightarrow \infty$ to be expected for nucleated polymerisation. However, as shown in Fig. \ref{fig2}, for intermediate times the mean polymer length overshoots and attains very large values. Also, Fig. \ref{fig2} shows that the mean polymer length for very large and very small values of recombination rate constant, $k_f^+=10^{8}, 10^{-8}$, $M s^{-1}$ the term $k_f^+ P^2$ has no effect on the early time kinetics for any reasonable value of $k_f^+$. This justifies our approximation of neglecting the term representing the contribution of polymer recombination as in this paper we are mainly interested in comparing the early time kinetics of \textit{end evaporation and addition} and \textit{body evaporation and addition}. The early time kinetics is characterised by the lag time, i.e., the time intercept of the tangent at the inflection point of the polymerisation curve, i.e., $M(t)$. \cite{hellstrand} For the sake of simplicity, we limit ourselves to strongly nucleated polymerisation, and set $k_n^+ \rightarrow 0$. This also reduces the degree of polynomial on the right hand side of Eqs. \ref{eq6} and \ref{eq7}, enabling us to obtain analytical solutions. In the limit $k_n^+ \rightarrow 0$, the polymerisation process has to be seeded, i.e., some initial polymerised mass, $M(0)=P(0) \neq 0$, has to be provided, otherwise the system stays in the initial state $M(0)=P(0)=0$.

The resulting dynamical equations become, after employing the approximations and rearranging terms in Eqs \ref{eq6} and \ref{eq7},
\bea \label{eq11}
\frac{dP(t)}{dt} &=& \kmf \left( M(t)-(2 n_c -1) P(t) \right),
\eea
for the polymer number concentration, and
\bea \label{eq12}
\frac{dM}{dt} &=& -M(t) (\alpha + \kpb M(t)) + P(t) (\beta - \gamma M(t)),
\eea
for the polymerised mass, where we introduce new dynamical constants $\alpha \equiv \kmb -\kpb \mtot$, $\beta \equiv 2(\kmb-\kme)+\kpb\mtot+2\kpe \mtot$ and $\gamma \equiv \kpb+2\kpe$. It should be emphasized that we define the parameters $\alpha$, $\beta$ and $\gamma$ merely for notational simplicity and do not associate any physical meaning to them. Eqs. \ref{eq11} and \ref{eq12} are nonlinear in nature, hence, the first trivial analysis demands the linearisation of these equations. Researchers have analysed similar equations for variety of pathways and obtained $t^2$ or $t^3$ time dependence for the polymerised mass fraction. \cite{ferron_early} However, for our particular choice of molecular pathways, the resulting linearised equations always yield simple exponential time dependence that in turn result into a linear time dependence of early times.	

\begin{figure}[!ht]
\begin{center}
\includegraphics[width=6.5in]{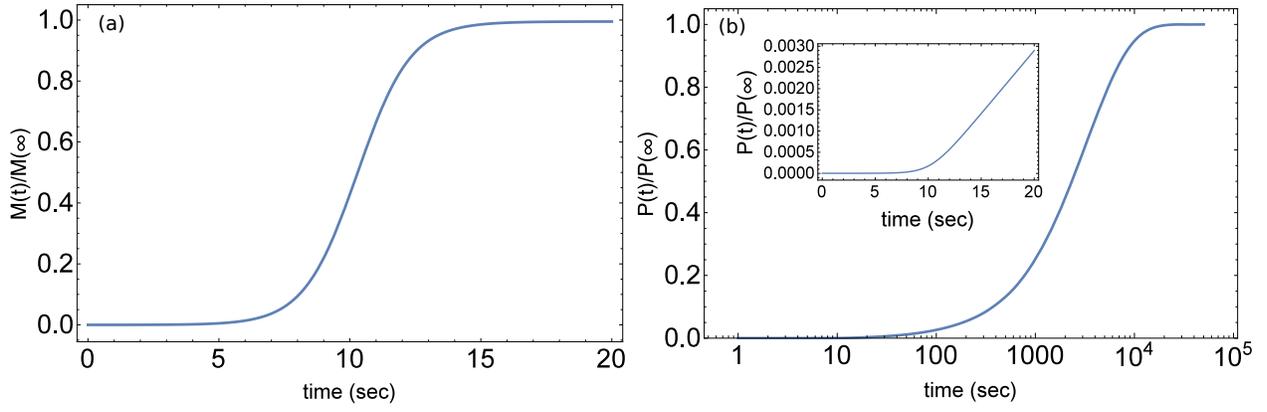}
\end{center}
\caption{Time evolution of the polymerised mass fraction $M(t)/M(\infty)$ and the polymer concentration $P(t)/P(\infty)$ obtained by numerically solving Eqs. \ref{eq6} and \ref{eq7} for $\kpf=0$ $M^{-1} s^{-1}$. The remainder of the system parameters are $k_n^+=10^{-5} s^{-1}$, $\kpe=10^{5} M^{-1} s^{-1}$, $\kme=10^{-2} s^{-1}$, $\kpb=10^{5} M^{-1} s^{-1}$, $\kmb=10^{-2} s^{-1}$, $\kmf=10^{-4} s^{-1}$, $\mtot=10^{-5} M$, $M(0)=\mtot \times 10^{-5}$ and $P(0)=\mtot \times 10^{-6}$, where, $M$ is moles per liter and $s$ is in seconds. The inset shows the early time behaviour of the number of polymers $P(t)$, which remains essentially constant during the relaxation of the polymerised monomeric mass, $M(t)$, and only relaxes after that. This figure highlights our claim of separation of time scales between the time evolution of the polymerised mass fraction and that of the polymer concentration. See the main text.}
\label{fig3}
\end{figure}

Although the dynamical equations Eqs. \ref{eq11} and \ref{eq12} contain the polymer scission term, we can reduce the effect of polymer scission on the kinetics of the polymerised monomeric mass, $M$, by choosing a very small value for the scission rate constant $\kmf$. The value of $\kmf=10^{-4}$ that we chose is small enough to diminish the effect of scission for the early time kinetics of the polymerised mass, as we find and also show below that this is true for our choice of set of parameters. For this small value of $\kmf$, the typical trajectory of the two moments is shown in Fig. \ref{fig3}. We notice that the polymerised mass $M$ evolves much faster in time than the number density of polymers $P$. The inset in Fig. \ref{fig3}b shows that in the asymptotic limit of vanishing scission rate $\kmf$, the lag phase for both moments is characterised by an approximately equal time scale. However, after the lag phase, $M$ relaxes to its equilibrium value much faster than $P$ does. Hence, for the early time kinetics of the polymerised monomeric mass, the number concentration of polymers remains effectively constant, i.e., equal to initial value $P(t)=P(0)$.

\begin{figure}[!ht]
\footnotesize
\stackunder[5pt]{\includegraphics[width=3.1in,height=2.35in]{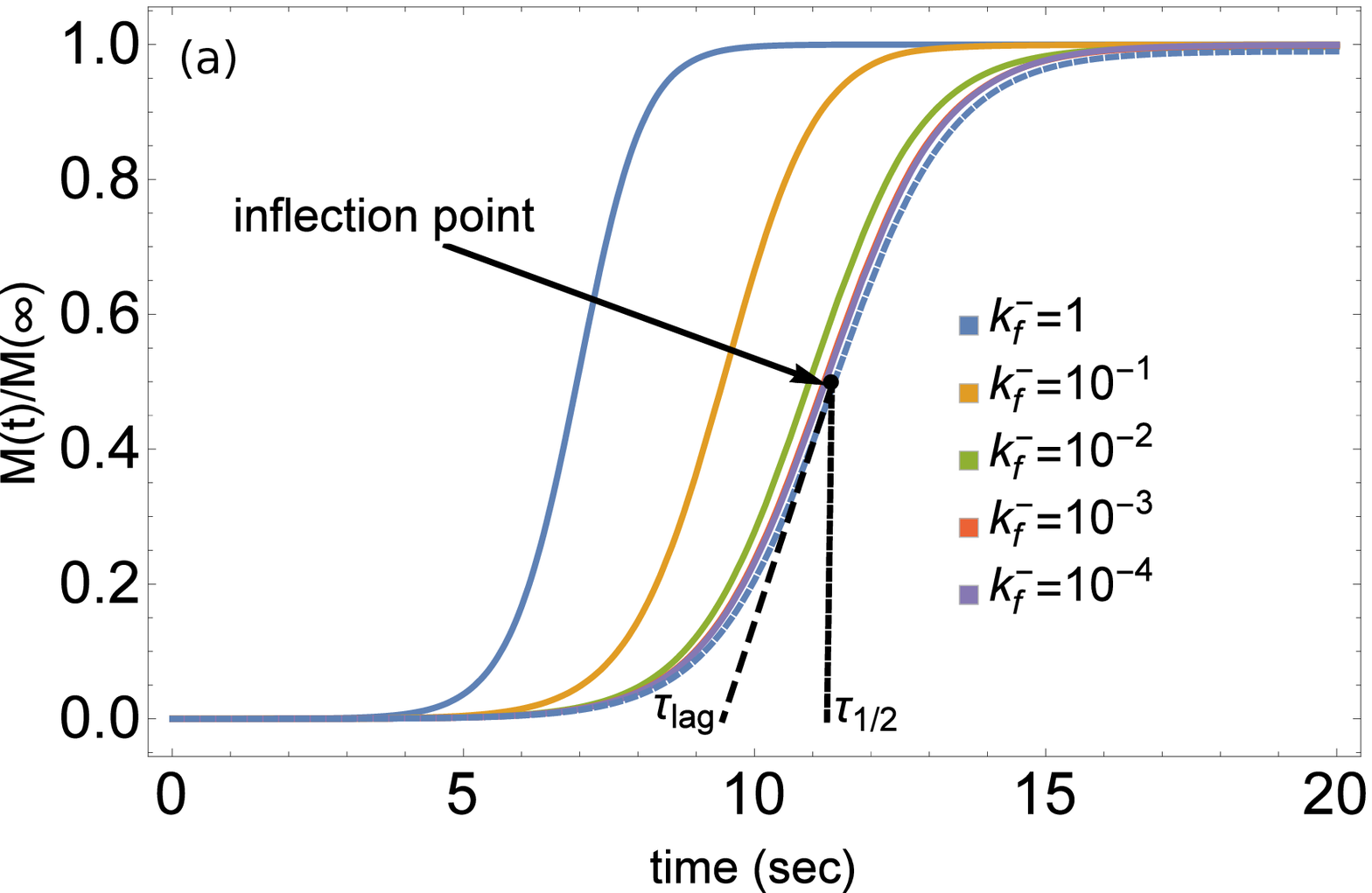}}{}%
\hspace{1cm}%
\stackunder[5pt]{\includegraphics[width=3.1in,height=2.35in]{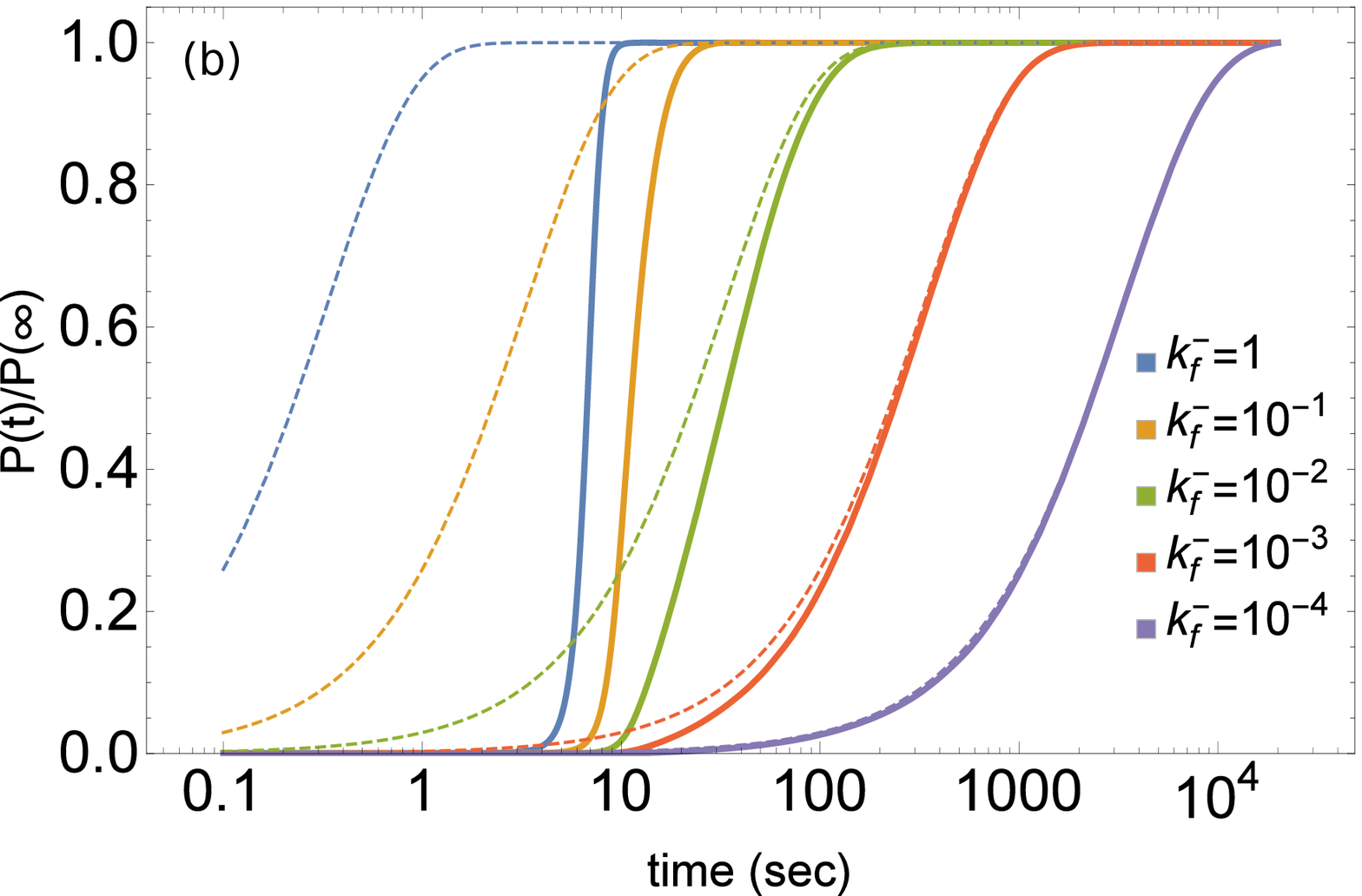}}{}
\caption{Time evolution of (a) the polymerised mass fraction $M(t)/M(\infty)$ and (b) the renormalized polymer concentration $P(t)/P(\infty)$ obtained by numerically solving Eqs. \ref{eq6} and \ref{eq7} (solid curves) for different scission rates $\kmf=1,10^{-1},10^{-2},10^{-3},10^{-4}$ and $\kpf=0$ $M^{-1} s^{-1}$. The dashed curves are the asymptotic analytical solutions Eqs. \ref{eq13} and \ref{eq14} for vanishing scission rate $\kmf$ in the limit of $k_n^+ \rightarrow 0$, i.e., strongly nucleated polymerisation. The remainder of the system parameters are same as used in Fig. \ref{fig3}.}
\label{fig4}
\end{figure}

Making use of this, we can solve Eq. \ref{eq11} for $M(t)$ for early times, yielding the explicit solution,
\bea \label{eq13}
\kpb M(t) &=& \frac{ \eta -\sigma}{2} - \frac{\eta \left(\eta-2 \kpb M(0)-\sigma \right)}{ \left(e^{t \eta} \left(\eta+2 \kpb M(0)+\sigma \right)+\eta-2 \kpb M(0)-\sigma \right)},
\eea
where for the purpose of notational simplicity we define $\sigma \equiv \alpha+\gamma P(0)$ and $\eta \equiv \sqrt{\sigma ^2+4 \beta  \kpb P(0)}$. In Fig. \ref{fig4} we show that Eq. \ref{eq13} agrees with the numerical results obtained in the limit $\kmf \rightarrow 0$. Below we will analyse the effect of molecular pathways by calculating the lag time from Eq. \ref{eq13}. For now, the main conclusion is that Eq. \ref{eq13} predicts sigmoidal kinetics, i.e., a lag phase followed by exponential growth and subsequent saturation of the solution for $M(t)$.

Next, we solve for $P(t)$ in the long-time limit when $M(t)$ has achieved its equilibrium value $M(\infty)$. For long times, i.e., post-lag phase of the polymerised mass fraction, $P(t)$ is given by
\bea \label{eq14}
\hspace{-25pt} P(t) &=& \frac{\left(e^{3 \kmf t}-1\right) \left[ M(\infty) \left(k_n^+ M(\infty)+\kmf-2 k_n^+ \mtot \right] + k_n^+ \mtot^2 \right) +3 \kmf P(0)}{3 \kmf e^{3 \kmf t}}.
\eea
Notice that for the analytical solution of the number concentration of polymers, the time scale of the evolution of $P(t)$ depends only on the scission rate constant $\kmf$. This is due to our choice of reaction scheme, where the \textit{scission and recombination} affects only the number of polymers. In contrast, \textit{end evaporation and addition} and \textit{body evaporation and addition} only affect the exchange of monomers between the free monomer pool and the polymer pool.

Having obtained a closed analytical expression for the time evolution of the polymerised mass, $M(t)$, we can now calculate the lag time, $\tlag$. This is achieved by calculating the inflection point or the point of maximum growth rate, i.e., the time at which the second derivative of Eq. \ref{eq13} is equal to zero. We then calculate the time intercept of the tangent at the inflection point, resulting in the analytical expression for the lag time. In general, for nucleated polymerisation kinetics the lag time $\tlag$ is a linear combination of two terms. \cite{hong_moment_closure} These terms represent two important characteristics of the polymerisation kinetics: the half-time $\thalf$ and apparent growth rate $\kapp$. The half-time $\thalf$ is the time at which the polymerised monomeric mass is exactly half of its steady-state (long-time) value and from Eq. \ref{eq13} we find that it is equal to the inflection point. Furthermore, the apparent growth rate $\kapp$ is the growth rate of the polymerisation curve at the inflection point (so, the time derivative of $M(t)$ at $t=\thalf$).

Hence, we find
\bea \label{eq15}
\tlag = \thalf - \frac{1}{2 \kapp},
\eea
where
\bea \label{eq16}
\thalf=\frac{\log \left(\frac{2 \eta }{\eta +2 \kpb M(0)+\sigma }-1\right)}{\eta},
\eea
denotes the half-time and
\bea \label{eq17}
\kapp = - \frac{\eta ^2}{2 \sigma},
\eea
the apparent growth rate.

To investigate the influence of the overall monomer concentration on the lag time, let us assume that all the other parameters, i.e., the rate constants, are constant and do not depend on the monomer concentration. This also ties in with our assumption that our polymerisation process is reaction-limited and not diffusion-limited. Let us first focus on the case $\kpb=\kpe$ and $\kmb=\kme$, i.e., which is true if \textit{body evaporation and addition} kinetics and \textit{end evaporation and addition} are equally likely. We see that $\alpha = \kmb -\kpb \mtot \approx -\kpb \mtot$ where we have $\kpb \mtot \gg \kmb$ for polymers to exist. The other parameters hence become $\beta = 2(\kmb-\kme)+\kpb\mtot+2\kpe \mtot \approx 3 \kpb \mtot$ and $\gamma = \kpb+2\kpe = 3\kpb \approx $ constant, i.e., independent of total monomer concentration. This further implies that $\sigma =\alpha + \gamma P(0) \approx -\kpb \mtot$, because $\alpha \sim \kpb \mtot \gg \gamma P(0) = 3 \kpb P(0)$ as $\mtot \gg P(0)$. Finally, the denominator in the expression for half-time is $\eta = \sqrt{\sigma^2 + 4 \beta \kpb P(0)} \approx \sqrt{(\kpb \mtot)^2+ 12 \kpb \mtot P(0)}$.

\begin{figure}[!ht]
\footnotesize
\stackunder[5pt]{\includegraphics[width=3.1in,height=2.35in]{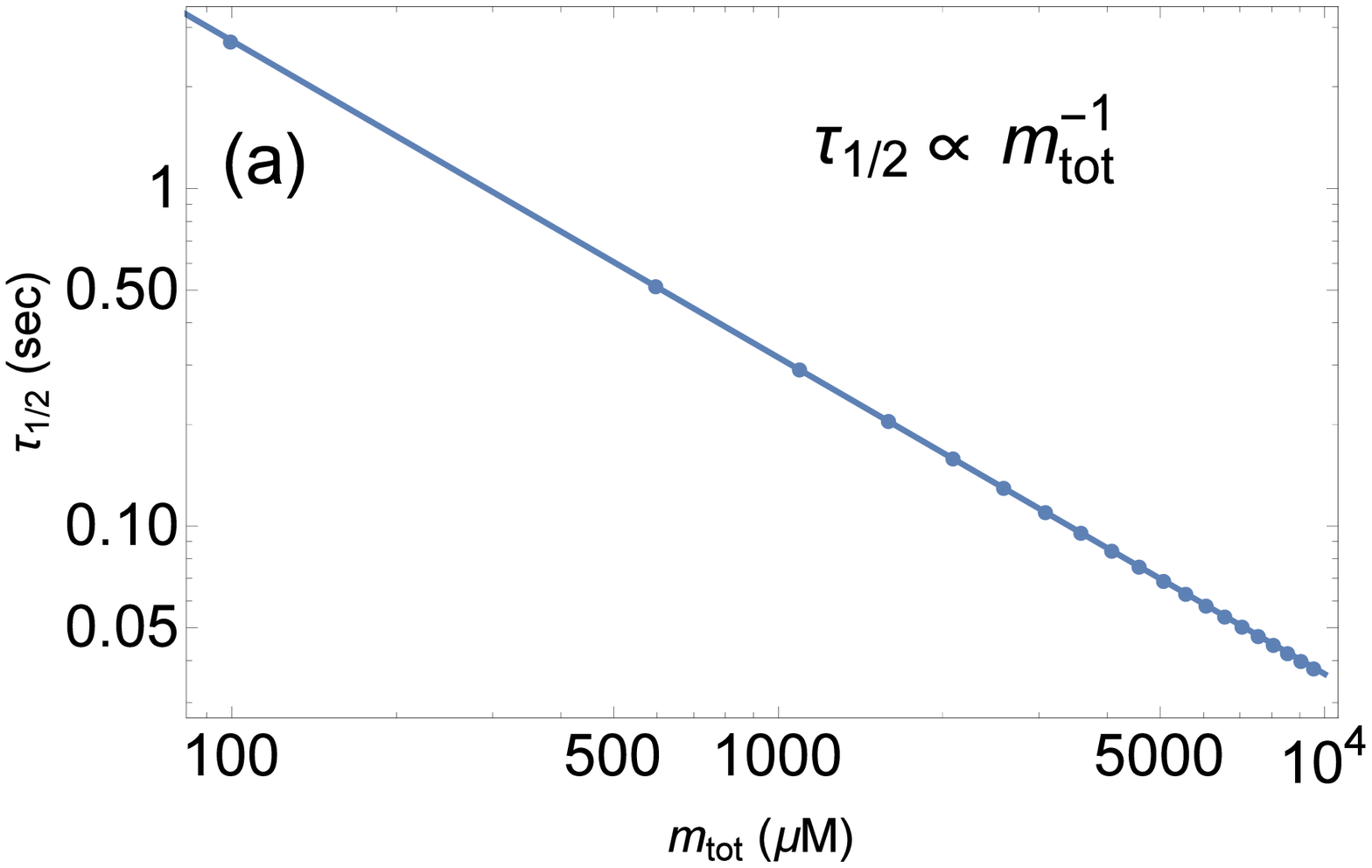}}{}%
\hspace{1cm}%
\stackunder[5pt]{\includegraphics[width=3.1in,height=2.35in]{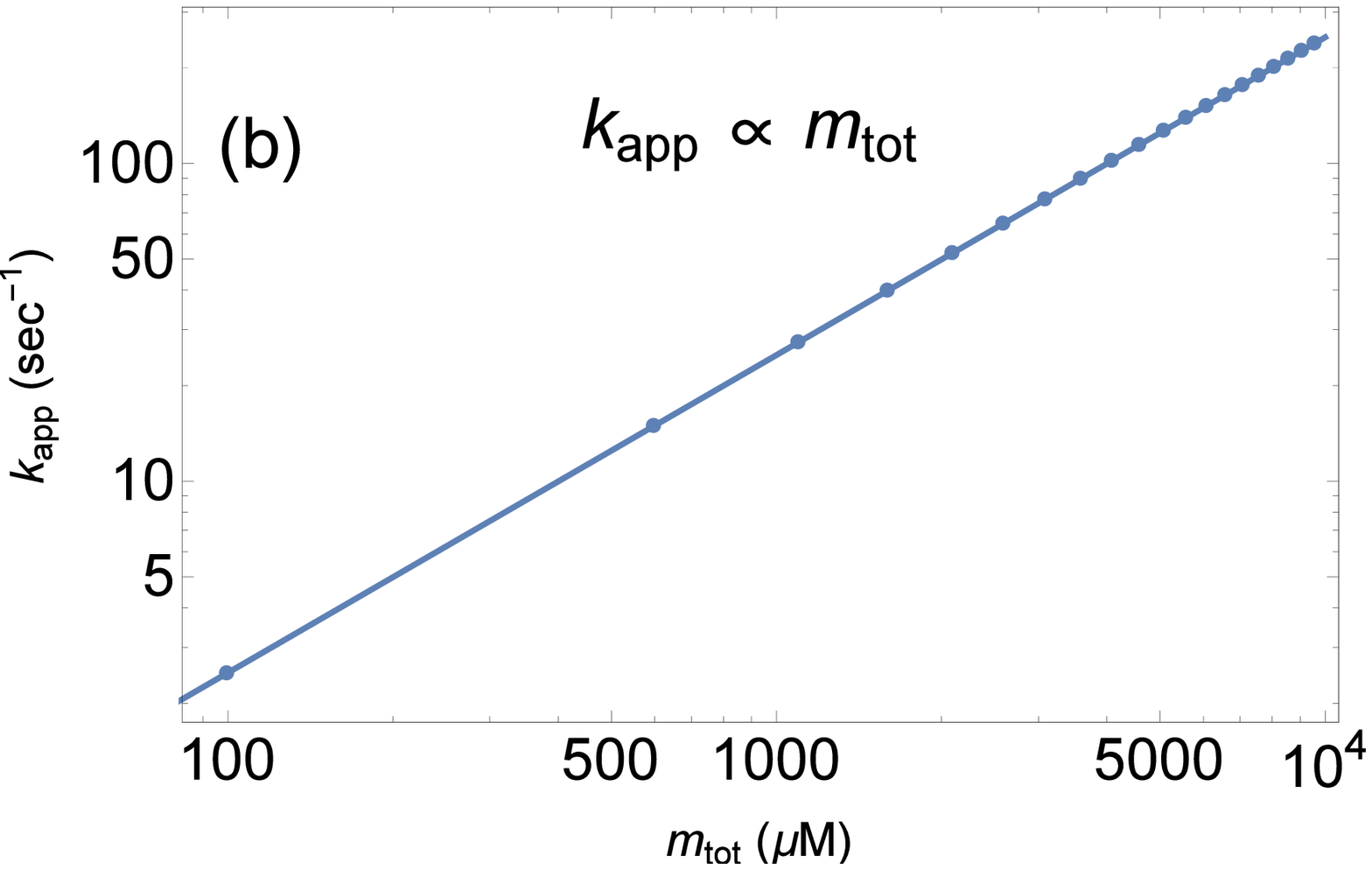}}{} 
\caption{(a) The half-time and (b) the apparent growth rate are shown on a double logarithmic scale as a function of the total monomer concentration for the case when \textit{bulk evaporation and addition} pathway is dominant. The half-time scales as an effective power law with the total monomer concentration, as is clear from the double logarithmic presentation. The circles are numerical value of the half-time and the apparent growth rate from Eqs. \ref{eq16} and \ref{eq17}, and the continuous line is a fit to that data, whose slope provides us with the power law exponent. The rest of the system parameters are $k_n^+=10^{-5} s^{-1}$, $\kpe=5 \times 10^{4} M^{-1} s^{-1}$, $\kme=10^{-2} s^{-1}$, $\kpb=5 \times 10^{4} M^{-1} s^{-1}$, $\kmb=10^{-2} s^{-1}$, $\kmf=10^{-4} s^{-1}$, $\mtot=10^{-5} M$, $M(0)=\mtot \times 10^{-4}$ and $P(0)=\mtot \times 10^{-6}$, where, $M$ is moles per liter and $s$ is in seconds.}
\label{fig5}
\end{figure}

Again, as $P(0) \ll \mtot$, we infer that $\eta \approx \kpb \mtot$. From Eq. \ref{eq16} the half-time has a logarithmic numerator resulting in a weak dependence of the numerator on $\mtot$, and effectively we have $\thalf \sim \eta^{-1} \sim \mtot^{-1}$. Similarly, $\kapp=\eta^2/2 \sigma$, where $\eta \approx \kpb \mtot$, and hence the apparent growth rate scales as $\kapp \sim \mtot$. In Fig. \ref{fig5}, we show, by fitting the concentration dependence of the half-time Eq. \ref{eq17} on a double logarithmic scale that our predictions for the power laws are indeed correct. It confirms that the logarithmic correction is indeed negligible. These power laws for the half-time and the apparent growth rate have same exponents as for the case of \textit{end evaporation and addition} in the limit of weak scission obtained before. \cite{hong}

\begin{figure}[!ht]
\footnotesize
\stackunder[5pt]{\includegraphics[width=3.1in,height=2.35in]{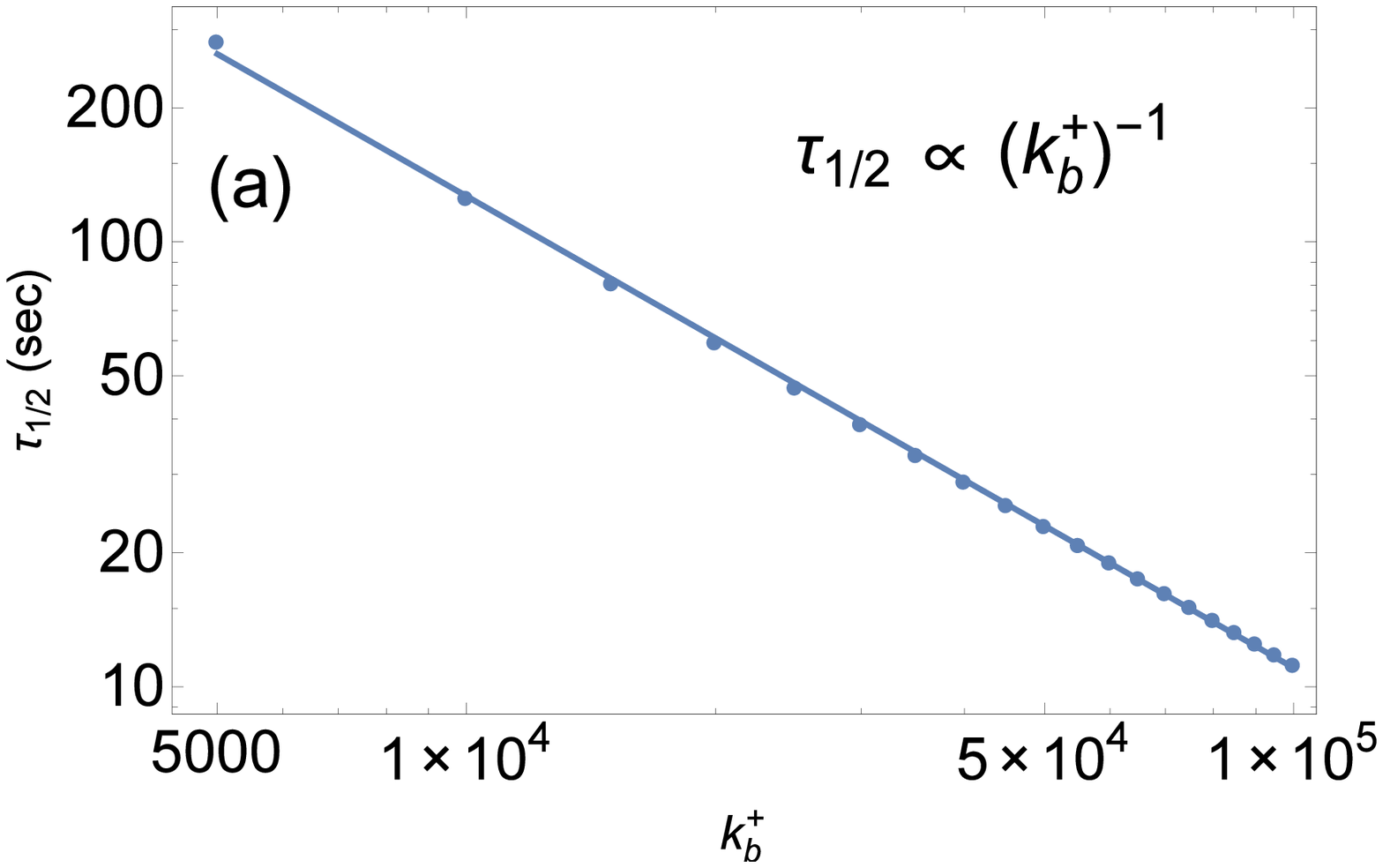}}{}%
\hspace{1cm}%
\stackunder[5pt]{\includegraphics[width=3.1in,height=2.35in]{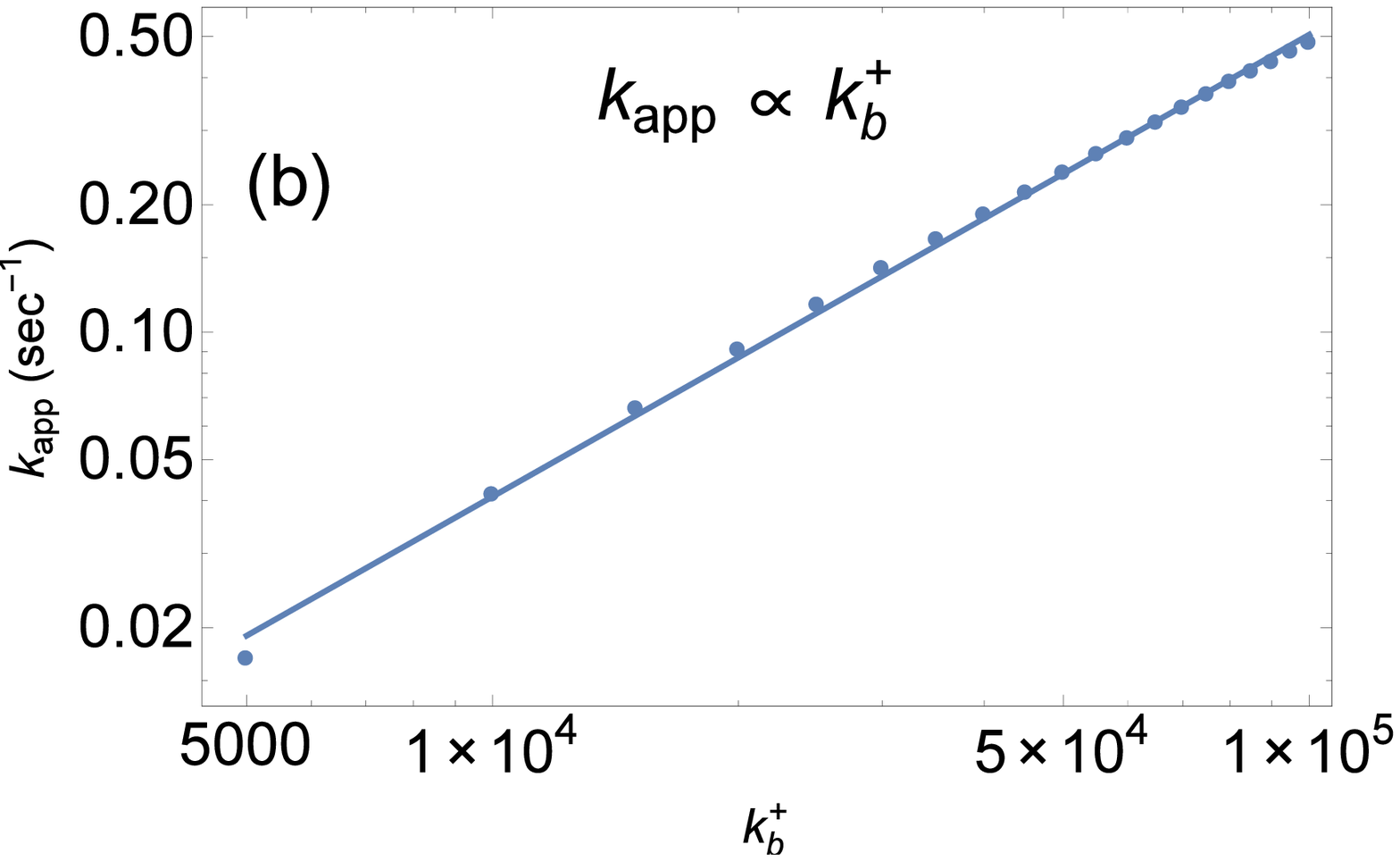}}{}
\caption{(a) The half-time and (b) the apparent growth rate are shown on a double logarithmic scale as a function of the forward rate constant of monomer insertion anywhere along the polymer back but the ends, in the absence of the \textit{end evaporation and addition} pathway. The circles are numerical values of the half-time and the apparent growth rate from Eqs. \ref{eq16} and \ref{eq17},and the continuous line is a fit to that data. The remainder of the system parameters are same as in Fig. \ref{fig5}}
\label{fig6}
\end{figure}

Let us now examine how the half-time and the apparent growth rate depend on the forward rates of the \textit{body evaporation and addition} and the \textit{end evaporation and addition} pathways, when only one of them is present. In the absence of \textit{end evaporation and addition} pathway, i.e., $\kpe=\kme=0$, we have $\eta \sim \kpb \mtot$, as shown above. This, in combination with the fact that the half-time scales as $\thalf \sim \eta^{-1}$ and that the apparent growth rate scales as $\kapp \sim \sigma/\eta^{-1}$, where $\sigma \sim - \kpb \mtot$, results in scaling of $\thalf \sim (\kpb)^{-1}$ and $\kapp \sim \kpb$. This is confirmed in  Fig. \ref{fig6}.


In the absence of \textit{body evaporation and addition} only primary nucleation and the \textit{end evaporation and addition} pathways are active, in the limit where \textit{scission and recombination} is sufficiently weak. In that limit, our equations are exactly the same as those presented in earlier works. \cite{hong} From the work of Hong et al., we already know that for weak scission, both the half time and the reciprocal apparent growth rate scale linearly with the total monomer concentration, which is the same as what we found for \textit{body evaporation and addition}. \cite{hong} In other words, from the monomer concentration dependence of the lag time we cannot distinguish body from end evaporation and addition. However, differences do show up when considering the dependence of the lag time on the forward rate constants. For the \textit{end evaporation and addition}, the half time and the reciprocal apparent growth rate scale as the square root of the forward rate constant of monomer addition, i.e., $\sqrt{\kpe}$, \cite{hong} which contrasts with what we found for \textit{body evaporation and addition}. For the latter we found a linear scaling with the forward rate constant.

In conclusion, although the scaling of the lag time  with the total monomer concentration is the same for both pathways, there is a difference in the scaling with the forward rate constants, or, equivalently, with the equilibrium constants for the two pathways. This in principle provides us with a probe to inspect the existence of \textit{body evaporation and addition} pathway in the polymerisation process, at least if we knew how to control them.

\section{Discussion and Conclusions} \label{sec4}
In our model calculations, we compare predictions based on the newly discovered kinetic pathway of \textit{body evaporation and addition} with the well studied pathway of \textit{end evaporation and addition}, in the context of strongly nucleated reversible polymerisation. To be able to do that, we rely on a kinetic scheme that includes a third pathway, known as \textit{scission and recombination}, that in the end we switch off asymptotically. \cite{cates_sr, tuomas_review} The presence of this third pathway is needed solely for mathematical reasons, to achieve our purpose. In practice, we cannot exclude the presence of \textit{scission and recombination}, and our set of equations, which focus on the two moments of the full distribution, allow for that.

The way the \textit{scission and recombination} and \textit{end evaporation and addition} pathways are implemented in the literature creates a conundrum, in which a process where a monomer is removed from the end can either be seen as a polymer scission or as an end evaporation event. \cite{hong, tuomas_review} To remedy this, we have implemented the \textit{scission and recombination} pathway for strongly nucleated linear reversible polymerisation in such a way that a polymer can break, resulting into the formation of two polymers of size greater than or equal to the \textit{critical nucleus}, which in our model has to be bigger than or equal to a dimer. We modelled polymer recombination, so the merging of two short polymers into a longer one the same way. Also we exclude the case $n_c=1$. This prescription prevents the \textit{scission and recombination} pathway from  interfering with the \textit{end evaporation and addition}. So in our prescription \textit{scission and recombination} does not directly affect the monomer pool.

In our view, our alternative implementation of the two pathways is the more sensible one, because detachment of a monomer is inherently different from the breaking of a polymer. Indeed, the latter process should strongly depend on the length of the polymer, whilst the former arguably should not provided the polymer is sufficiently long. \cite{tom_length_dependence} In addition, our implementation of the \textit{scission and recombination} pathway reproduces the thermodynamically consistent law of mass action for the amount of polymerised material under conditions of equilibrium and in the presence of the \textit{end evaporation and addition} pathway. The amount of polymerised material of course is only one of the moments of the full length distribution.
As for the other moments, such as the number of polymers in solution, they suffer from thermodynamically inconsistent long-time behaviour, a drawback that we share with previous studies. \cite{cohen, hong, tuomas_science}

However, if we focus on the short-time behaviour of the reversible polymerisation reactions, we could argue, as in fact is tacitly done in the literature, that the long-time behaviour of the system is inconsequential. \cite{hong, tuomas_review} Indeed, many thermodynamically inconsistent pathways have been shown to be in quantitative agreement with experimental findings regarding the early time kinetics, which is the prime focus of almost all experimental works. \cite{tuomas_science, knowles_two_stage, hong, hong_moment_closure, subramaniam} For this reason, we also focus on the early time kinetics, when comparing body and end evaporation and addition, in the limit of asymptotically weak scission. Because polymer recombination has been shown theoretically not to substantially alter the early-time polymerisation kinetics, we suppress the recombination step completely for mathematical expedience in order to obtain closed-form solutions, in the final steps of our analysis. \cite{knowles_two_stage}

\begin{figure}[!ht]
\footnotesize
\stackunder[5pt]{\includegraphics[width=3.1in,height=2.35in]{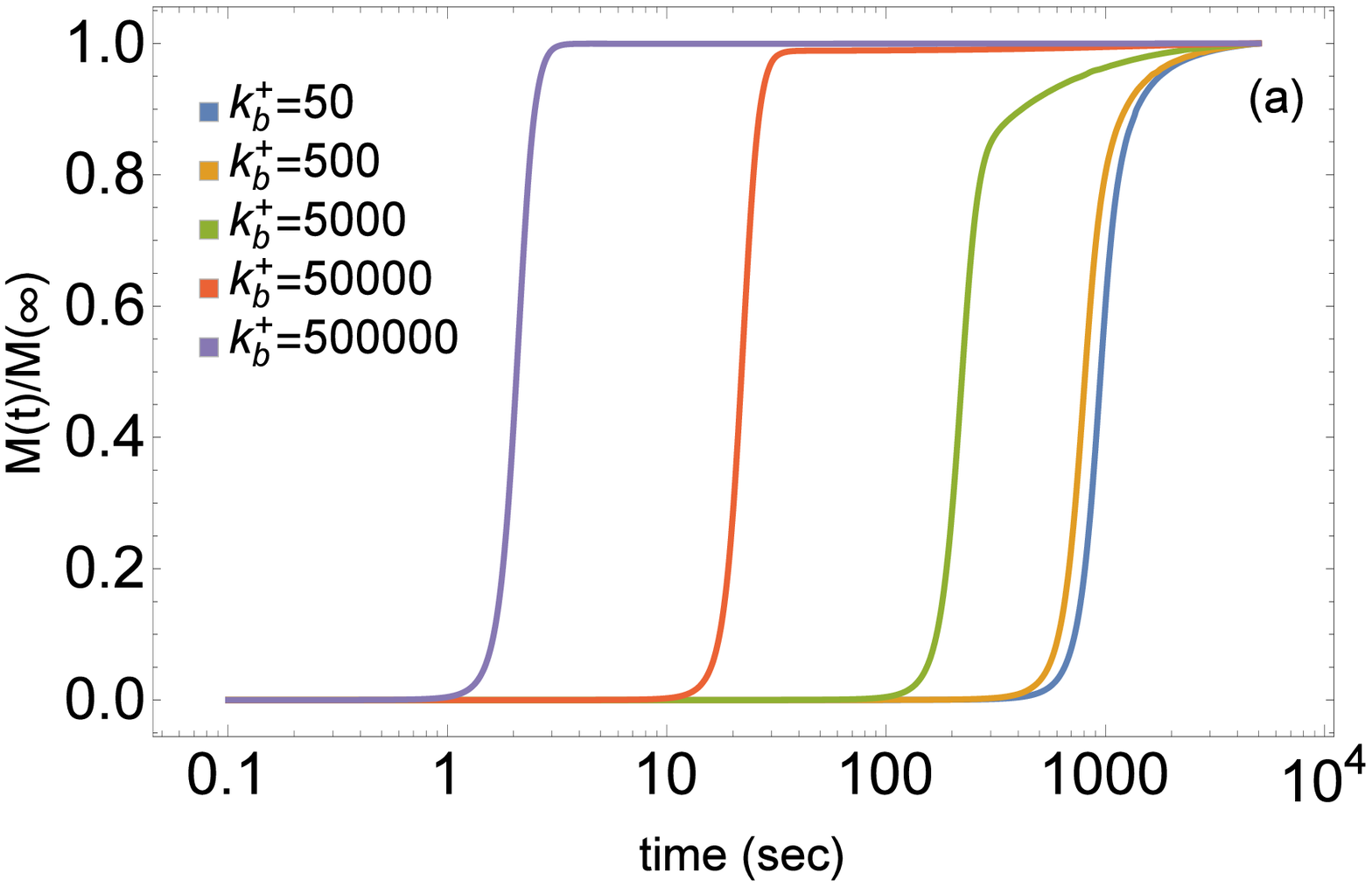}}{}%
\hspace{1cm}%
\stackunder[5pt]{\includegraphics[width=3.1in,height=2.35in]{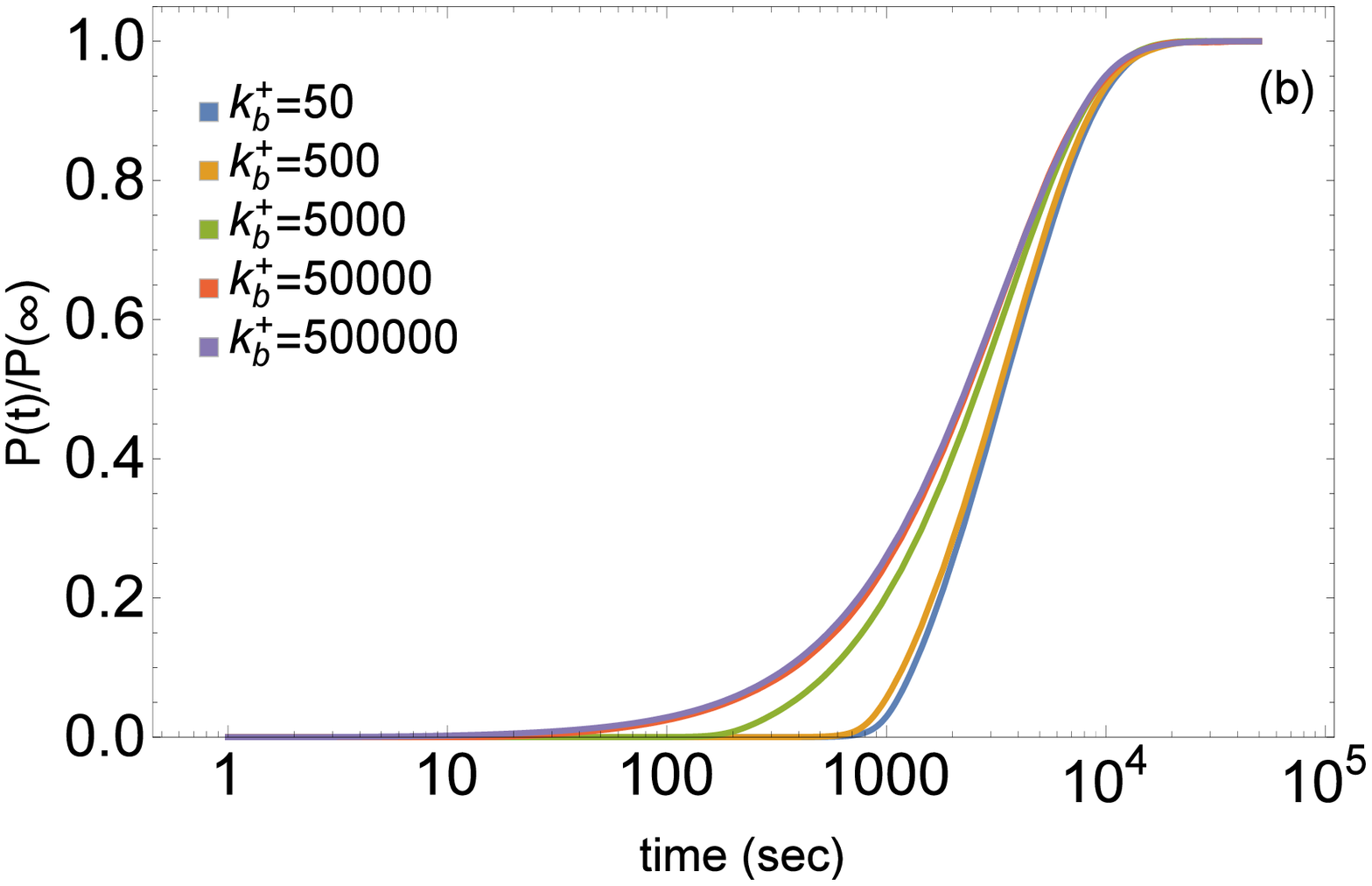}}{}
\caption{Time evolution of (a) the polymerised mass fraction $M(t)/M(\infty)$ and (b) the renormalized polymer concentration $P(t)/P(\infty)$ obtained by numerically solving Eqs. \ref{eq6} and \ref{eq7} for different scission rates $\kpb=5\times10,5\times10^{2},5\times10^{3},5\times10^{4},5\times10^{5}$ $M^{-1} s^{-1}$ and $\kpf=0$ $M^{-1} s^{-1}$. The remainder of the system parameters are same as used in Fig. \ref{fig3}.}
\label{fig7}
\end{figure}

In this limit, we find that the typical polymerisation kinetics of \textit{body evaporation and addition} is characterised by a separation of time scales between the polymerised monomeric mass $M$ and the polymer number concentration $P$. This is obvious from the difference in the lag times for these two quantities, i.e., the time scales required to get a substantial growth. Such separation of time scales has also been found for the \textit{end evaporation and addition} pathway in combination with weak polymer scission and no recombination albeit that it is  is not as strong. \cite{hong_moment_closure} To illustrate this, in Fig. \ref{fig7}, we show the polymerised mass fraction and the polymer concentration as a function of time, where we vary the forward rate of body addition at a fixed value of the forward rate of end addition. We cover the full spectrum from predominant body to predominant end evaporation and addition.

Comparing Fig. 7a and 7b, it is clear that the separation of time scales is many orders of magnitude larger for \textit{body evaporation and addition} than that for \textit{end evaporation and addition}.
Indeed, the polymerised mass fraction evolves much faster in time than the number of polymers does, if the \textit{body evaporation and addition} pathway is active. It is this vast separation of time scales that allows us to obtain first order perturbative solutions to the two moment equations that are in quantitative agreement with numerical results for vanishing scission rate constant.

Our perturbative solutions for the polymerised mass fraction and the polymer concentration, are sigmoidal as a function of time and provide us with the closed form solution for the lag time. The lag time associated with the polymerised mass fraction is the one that is experimentally readily accessible, and hence we focus on that. We find, within the limit where it is valid, that the lag time produces only a weak dependence on the initial polymerised monomeric mass. This contrasts with other studies where they do not include \textit{body evaporation and addition} in the reaction scheme, and is a result of the extremely fast kinetics connected with that pathway. \cite{oosawa, cohen, hong, hong_moment_closure}

In the usual definition of the lag time, it takes the form of the sum of a half-time $\thalf$ and a reciprocal of an apparent growth rate $\kapp$. \cite{tuomas_review} For \textit{body evaporation and addition} we find the same linear scaling of the the half time and the reciprocal apparent growth rate with the total monomer concentration, $\mtot$, as was found in the literature for the \textit{end evaporation and addition}. \cite{hong} In fact, for the \textit{bulk evaporation and addition} pathway it is the product $\kpb \mtot$ dictates the scaling of the lag time, implying that both time scales are also linearly dependent on the forward rate for the body addition, $\kpb$. In contrast, for the \textit{end evaporation and addition} pathway both time scale are proportional to the square root of the forward rate constant of the end addition, $\sqrt{\kpe}$. \cite{hong}

It seems that the newly proposed pathway, although it speeds up the growth of the polymerised mass by providing a larger number of places to insert or remove a monomer from a polymer, does not affect the dependence of the lag time on the total monomer concentration. However, it does affect how the relevant forward rate constant influences that time scale. This means that to distinguish end from body evaporation and addition experimentally by probing the dependence of the lag time on the system parameters, we would need to be able to control this quantity. This is not trivial. It may well be that highly specialised techniques, such as those used in reference \cite{albertazzi}, are required to actually observe it. In fact, this may even be the reason, why it had not been considered before.

\section{Acknowledgements}
We thank Mariana Oshima Menegon, Shari Finner and Stefan Paquay for a critical reading of the manuscript. This work was supported by the Nederlandse Organisatie voor Wetenschappelijk Onderzoek through Project No. 712.012.007.

\appendix
\section{Derivation of Moment Equations from Discrete Master Equation}
We start by writing down the discrete master equation, Eq. \ref{eq2}, for the reaction schemes described in the main text. Next, we define the principal moments, the number of polymers $P$ and the polymerised monomeric mass $M$, Eqs. (7) and (8).

After substituting Eq. \ref{eq4} in Eq. \ref{eq5}, the equation for the number of polymers $P$ is given by,
\bea
\frac{dP(t)}{dt} &=& \sum_{i=n_c}^{\infty} k_n^+ x(t)^{n_c} \delta_{i,n_c} + 2 \kpe \sum_{i=n_c}^{\infty} x(t) y_{i-1}(t) - 2 \kpe x(t) \sum_{i=n_c}^{\infty} y_{i}(t) + 2 \kme \sum_{i=n_c}^{\infty} y_{i+1}(t) - 2 \kme \sum_{i=n_c}^{\infty} y_i(t) \nonumber \\
& & + \sum_{i=n_c}^{\infty} \kpb (i-2) y_{i-1}(t) - \sum_{i=n_c}^{\infty} \kpb (i-1) x(t) y_{i}(t) + \sum_{i=n_c}^{\infty} \kmb (i-1) y_{i+1}(t) - \sum_{i=n_c}^{\infty} \kmb (i-2) y_i(t) \nonumber \\
& & - \sum_{i=n_c}^{\infty} \kmf (i- 2 n_c +1) y_i(t) + 2 \kmf \sum_{i=n_c}^{\infty} \sum_{j=i+n_c}^{\infty} y_j(t) \nonumber \\
& & + \kpf \sum_{i=n_c}^{\infty} \sum_{k+l=i} y_k(t) y_l(t) - 2 \kpf \sum_{i=n_c}^{\infty} y_i (t) \sum_{j=n_c}^{\infty} y_j.
\eea
To obtain a kinetic equation without any summation signs, we collect and simplify the equation term by term. Let us start with the terms accounting for monomer addition at the end,
\bea
& & \left ( \sum_{i=n_c}^{\infty} y_{i-1} - \sum_{i=n_c}^{\infty} y_{i} \right) = \left ( \sum_{i=n_c-1}^{\infty} y_{i} - \sum_{i=n_c}^{\infty} y_{i} \right) = y_{n_c-1} = 0.
\eea
Notice that $y_{n_c-1}=0$, because the smallest stable polymer is the critical nucleus of size $n_c$. Similarly for end evaporation,
\bea
& & \left ( \sum_{i=n_c}^{\infty} y_{i+1} - \sum_{i=n_c}^{\infty} y_{i} \right) = \left ( \sum_{i=n_c+1}^{\infty} y_{i} - \sum_{i=n_c}^{\infty} y_{i} \right) = - y_{n_c}.
\eea
The analysis in the main text assumes the strongly nucleated polymerisation, i.e., $k_n^+ \rightarrow 0$. In this limit the critical nuclei are highly unstable and can be neglected.

For terms six and seven in Eq. (A1) that represent monomer addition on the polymer backbone, we obtain
\bea
\kpb \sum_{i=n_c}^{\infty} \left( (i-2) y_{i-1} - (i-1) y_i \right) = \kpb \sum_{i=n_c-1}^{\infty} (i-1) y_i - \kpb \sum_{i=n_c}^{\infty} (i-1) y_i = 0.
\eea
Similarly, terms eight and nine in Eq. (A1) accounting for monomer removal from the polymer backbone, simplify to
\bea
\sum_{i=n_c}^{\infty} (i-1) y_{i+1} - \sum_{i=n_c}^{\infty} (i-2) y_i = \sum_{i=n_c}^{\infty} (i-1) y_{i+1} - \sum_{i=n_c-1}^{\infty} (i-1) y_i = - (n_c-1) y_{n_c}.
\eea
Once again under the assumption of strongly nucleated polymerisation, we neglect $y_{n_c}$.

The contribution from polymer scission can be rewritten in terms of a theta function as
\bea
-  \sum_{i=n_c}^{\infty} (i- 2 n_c +1) y_i + 2  \sum_{i=n_c}^{\infty} \sum_{j=n_c}^{\infty} y_j \Theta (i-j-n_c) \nonumber \\
-  \sum_{i=n_c}^{\infty} (i- 2 n_c +1) y_i + 2  \sum_{i=n_c}^{\infty} (1+j-2 n_c) y_j =  (M-(2 n_c -1) P),
\eea
and the contribution of polymeric recombination is
\bea
\sum_{k=n_c}^{\infty} \sum_{l=n_c}^{\infty} y_k y_l - 2 \sum_{i=n_c}^{\infty} y_i \sum_{j=n_c}^{\infty} y_j = \left( \sum_{k=n_c}^{\infty} y_k \right) \left( \sum_{l=n_c}^{\infty} y_l \right) - 2 \left( \sum_{i=n_c}^{\infty} y_i \right) \left( \sum_{j=n_c}^{\infty} y_j \right) = -P^2
\eea
This gives us following equation for the number of polymers $P$,
\bea
\frac{dP(t)}{dt} = - \kpf P(t)^2 + \kmf \left( M(t)-(2 n_c -1) P(t) \right) + k_n^+ x(t)^{n_c}
\eea
For the polymerised monomeric mass $M(t)$, the evolution equation is,
\bea
\frac{dM(t)}{dt} &=& \sum_{i=n_c}^{\infty} k_n^+ i x(t)^{n_c} \delta_{i,n_c} + 2 \kpe x(t) \sum_{i=n_c}^{\infty} i y_{i-1}(t) - 2 \kpe x(t) \sum_{i=n_c}^{\infty} i y_{i}(t) + 2 \kme \sum_{i=n_c}^{\infty} i y_{i+1}(t) - 2 \kme \sum_{i=n_c}^{\infty} i y_i(t) \nonumber \\
& & + \sum_{i=n_c}^{\infty} \kpb i (i-2) y_{i-1}(t) - \sum_{i=n_c}^{\infty} \kpb i (i-1) x(t) y_{i}(t) + \sum_{i=n_c}^{\infty} \kmb i (i-1) y_{i+1}(t) - \sum_{i=n_c}^{\infty} \kmb i (i-2) y_i(t) \nonumber \\
& & - \sum_{i=n_c}^{\infty} \kmf i (i- 2 n_c +1) y_i(t) + 2 \kmf \sum_{i=n_c}^{\infty} \sum_{j=i+n_c}^{\infty} i y_j(t) + \kpf \sum_{i=n_c}^{\infty} \sum_{k+l=i} i y_k(t) y_l(t) \nonumber \\
& & - 2 \kpf \sum_{i=n_c}^{\infty} i y_i (t) \sum_{j=n_c}^{\infty} y_j(t).
\eea
The terms involving monomer addition at the ends are
\bea
2  x(t) \sum_{i=n_c}^{\infty} i \left( y_{i-1}(t)-y_i(t) \right) = 2  \left( \sum_{i=n_c-1}^{\infty} (i+1) y_i - \sum_{i=n_c}^{\infty} i y_i \right) = 2  x(t) P(t),
\eea
and the term arising from monomer removal at the ends are
\bea
2  \sum_{i=n_c}^{\infty} i \left( y_{i+1}(t)-y_i(t) \right) = 2  \left( \sum_{i=n_c+1}^{\infty} (i-1) y_i - \sum_{i=n_c}^{\infty} i y_i \right) = 2  (-P(t) - n_c y_{n_c}).
\eea
Where we neglect the contribution, $n_c y_{n_c}$.

Next the terms contributed by the monomer addition in the bulk becomes
\bea
& &  \sum_{i=n_c}^{\infty} \left( i (i-2) y_{i-1} - i (i-1) y_i \right) \nonumber \\
&=&  \left( \sum_{i=n_c-1}^{\infty} (i+1)(i-1) y_i - \sum_{i=n_c}^{\infty} i (i-1) y_i \right)\nonumber \\
&=&  \left( \sum_{i=n_c}^{\infty} (i^2-1) y_i + \sum_{i=n_c}^{\infty} i y_i - \sum_{i=n_c}^{\infty} i^2 y_i \right) \nonumber \\
&=&  x(t) (M(t) - P(t)),
\eea
and the terms from monomer removal from the bulk are
\bea
& &  \left( \sum_{i=n_c}^{\infty} i (i-1) y_{i+1} - \sum_{i=n_c}^{\infty} i (i-2) y_i \right) \nonumber \\
&=&  \left( \sum_{i=n_c+1}^{\infty} (i-1) (i-2) y_i - \sum_{i=n_c}^{\infty} i (i-2) y_i \right) \nonumber \\
&=&  - \sum_{i=n_c+1}^{\infty} (i-2) y_i =  (- n_c y_{n_c} + 2 P(t)- M(t)).
\eea
This completes our closure of discrete master equation to obtain moment equations, that are given by,
\bea
\frac{dP(t)}{dt} &=& - \kpf P(t)^2 + \kmf \left( M(t)-(2 n_c -1) P(t) \right) + k_n^+ x(t)^{n_c},
\eea
and
\bea
\frac{dM}{dt} &=& 2 \big( x(t) \kpe P(t) - \kme P(t) \big) + \kpb x(t) (M(t)+P(t)) + \kmb (2 P(t)-M(t))+ n_c k_n^+ x(t)^{n_c}.
\eea

\end{document}